\documentclass{ws-ijmpb}

\begin{document}

\markboth{Stefano Longhi}
{Non-Hermitian interaction of a discrete state with a continuum}

%
\catchline{}{}{}{}{}
%

\title{Non-Hermitian interaction of a discrete state with a continuum}

\author{STEFANO LONGHI}

\address{Dipartimento di Fisica e Istituto di Fotonica e Nanotecnologie del Consiglio Nazionale delle Ricerche, Politecnico di Milano, Piazza Leonardo da Vinci 32\\
Milan, 20133, Italy\\
stefano.longhi@fisi.polimi.it}

\maketitle


\begin{abstract}
The interaction of a discrete state coupled to a continuum is a longstanding problem of major interest in different areas of quantum and classical physics. 
In Hermitian models, several dynamical decoupling schemes have been suggested, in which the discrete-continuum interaction can be substantially reduced and even suppressed. In this work we consider a discrete state interacting with a continuum via a time-dependent {\em non-Hermitian} coupling with finite (albeit arbitrarily long) duration, and show rather generally that for a wide class of coupling temporal shapes, in which the real and imaginary parts of the coupling are related each other by a Hilbert transform, the discrete state returns to its initial condition after the interaction with the continuum, while the continuum keeps trace of the interaction. Such a behavior, which does not have any counterpart in Hermitian dynamics, can be referred to as non-Hermitian {\em pseudo} decoupling. Non-Hermitian pseudo decoupling is illustrated by considering a non-Hermitian extension of the Fano-Anderson model in a one-dimensional tight-binding lattice. Scuh a non-Hermitian model can describe, for example, photonic hopping dynamics in a tight-binding chain of optical microrings or resonators, in which non-Hermitian coupling can be realized by fast modulation of the real and imaginary (gain/loss) parts of the refractive index of the  edge microring. 
\end{abstract}


\section{Introduction}
The coupling of a discrete (bound) state with a continuum is a fundamental process found in different areas of physics, ranging from quantum mechanics \cite{r1,r1bis,r2,r3} to statistical physics \cite{r3bisA,r3bisB,r3bisC,r4,r5,r5bis,r5tris}, atomic and molecular physics \cite{r6,r6bis,r6trisA,r6trisB}, optics \cite{r7A,r7B,r8A,r8B}, quantum thermodynamics \cite{r8bis} and cosmology \cite{r9A,r9B,r9C} to mention a few. In quantum physics, discrete-continuum coupling is responsible for quantum mechanical decay and decoherence, whereas in optics it is responsible for important effects such as radiation losses in integrated photonic circuits. The ability of controlling and even suppressing the discrete-continuum coupling has attracted great interest for both fundamental aspects and practical applications. For example, in quantum physics it is known that the decay of a metastable state can be slowed down (Zeno effect)  or accelerated (anti-Zeno effect) by frequent observations of the system \cite{Z1A,Z1B,Z2,Z3,Z3bisA,Z3bisB,Z4,Z5A,Z5B}. Other dynamical methods, which do not require observation of the system, include the "bang-bang" control scheme, based on very rapidly and strongly driving pulses \cite{Z9A,Z9B,Z9C,Z9D}, and the periodic modulation of the coupling to the continuum \cite{Z10,Z11}. Dynamical decoupling methods can be applied to control macroscopic quantum tunneling  \cite{Z11,Z12} 
 and to tailor evanescent coupling of light in coupled optical structures \cite{Z13}. Quantum Zeno and anti-Zeno effects and their classical analogues have been studied and observed in many different physical setups such as cold atoms and Bose-Einstein condensates \cite{Z14,Z15A,Z15B,Z16}, superconducting qubits \cite{Z17}, spin systems \cite{Z18}, nuclear magnetic resonance systems \cite{Z19}, classical optical systems \cite{Z20A,Z20B,Z20C,Z20bisA,Z20bisB,Z20bisC} and cavity quantum electrodynamics \cite{Z21A,Z21B,Z21C}.\par 
 In such previous studies, the problem of discrete-continuum decoupling has been mostly considered in the framework of Hermitian systems, where the Hamiltonian $\hat{H}$ of the full system is an Hermitian operator.  However, a rapidly growing interest is currently devoted to study the dynamics of non-Hermitian quantum and classical systems \cite{r21,r21bis}, especially those possessing parity-time ($\mathcal{PT})$ symmetry \cite{r22}. Non-Hermitian Hamiltonians are generally introduced as effective models to describe open systems, where energy (particles) can be feed into (or extracted from) the system. For example, in models of inelastic nuclear scattering
theory \cite{referee1,referee2} nuclear absorption is phenomenologically  included in the analysis by the introduction of a fictitious imaginary scattering potential, the so-called optical potential, whereas imaginary potentials that act as source and  sink for carriers are introduced in non-Hermitian models of mesoscopic quantum transport \cite{referee3A,referee3B,referee4}. Similarly, imaginary potentials are used in non-Hermitian extensions of Bose-Hubbard models and of the Gross-Pitaevskii equation to describe the mean-field dynamics of a Bose-Einstein condensate with particle loss and particle gain \cite{referee5A,referee5B,referee5C,referee5D}. In optics, light propagation in dielectric media with spatial regions of optical loss and gain are described by non-Hermitian Hamiltonians \cite{referee6A,referee6B}. Recent works investigated theoretically the dynamics of a discrete state coupled to a continuum in the framework of non-Hermitian Hamiltonian models \cite{R1A,R1B,R1C,R1D,R1E,R2,R3}, predicting the existence of bound states either outside or embedded into the continuum \cite{R1A,R1B,R1C,R1D,R1E} and the suppression of decay for a bound state with an energy embedded in an unstable continuum \cite{R3}.  \par
In this work we disclose some peculiar features of non-Hermitian dynamics that arise when a discrete state interacts for a finite time with a continuum of states via a {\em non-Hermitian} coupling,  and predict a novel effect which is impossible to observe in the ordinary Hermitian dynamics and that we call {\em non-Hermitian pseudo-decoupling}.  Non-Hermitian couplings have been introduced in a few foundational or effective Hamiltonian models in different areas of physics \cite{r22,E1,E2,E3,E4}. In particular, non-Hermitian couplings have been used to describe a $\mathcal{PT}$-symmetric extension of the Lee model  in the ghost regime \cite{r22,E1,E2}, non-Hermitian Bose-Hubbard Hamiltonians \cite{E3}, and $\mathcal{PT}$-symmetric extensions of the Jaynes-Cumming model \cite{E4}. Effective models that yield non-Hermitian couplings are also encountered in the theory of the inverted quantum oscillators and quantum amplifiers \cite{E5,E6} and introduced in several other systems \cite{E7A,E7B,E7C,E7D,E7E,E7F,E7G,E7H}.  In particular, engineered optical structures where photonic transport arises from evanescent field coupling can be tailored to simulate an effective non-Hermitian interaction between a discrete state (such as an optical resonator or waveguide) with a continuum of states (such as a chain of optical resonators, a waveguide array, a slab waveguide or a photonic crystal)  \cite{ruffa,ruffa1}. Here we consider a time-varying non-Hermitian coupling of a discrete state with a continuum of states and show that for a wide class of coupling temporal shapes, in which the real and imaginary parts are related each other by a Hilbert transform (Kramers-Kronig relations), the discrete state returns to its initial condition after the interaction, while some excitation is left in the continuum thus keeping memory of the interaction. Such a phenomenon, which does not have any counterpart in a norm-conserving Hermitian system, can be referred to as {\em non-Hermitian  pseudo decoupling}, since the interaction is fully invisible for the discrete state, i.e. it is {\em effectively}  decoupled form the continuum as if the interaction had not occured, but not for the continuum, for which the interaction changes its initial state. The effect is illustrated by considering a paradigmatic model of a discrete state coupled to a tight-binding continuum, namely photonic transport in a non-Hermitian extension of the Fano-Anderson model. This model can be physically realized by a semi-infinite tight-binding optical lattice with on-site oscillating gain/loss at the edge site of the lattice.\par
The paper is organized as follows. In Sec.2 the model describing the dynamics of a discrete state coupled to a continuum via a time-dependent non-Hermitian coupling is introducedl, and an extension of the Fermi golden rule in the weak-coupling (Markovian) limit is derived, indicating the possibility of observing an effective pseudo decoupling whenever the non-Hermitian interaction has a vanishing effective interaction time.  Pseudo decoupling is rigorously proven in Sec.3 beyond the weak-coupling approximation and  illustrated for photonic transport in a tight-binding realization of the non-Hermitian Fano-Anderson model in Sec.4. Finally, the main conclusions are outlined in Sec.5. 

\section{Coupling of a discrete state with a continuum via a time-dependent non-Hermitian interaction}
\subsection{The model}
Let us consider the interaction of a discrete (normalizable) state $|a \rangle$ with a continuum $| \omega \rangle$ of states [Fig.1(a)], which is described rather generally by the Fano-Anderson Hamiltonian
\begin{eqnarray}
\hat{H} & = &  \omega_a | a \rangle \langle a | + \int d \omega   \omega | \omega \rangle \langle \omega | \\
&+ & f(t) \int \left\{ g(\omega) | a \rangle \langle \omega | + g^*( \omega) | \omega \rangle \langle a | \right\} \nonumber
\end{eqnarray}
where $\omega_a$ and $\omega$ are the (real) bare frequencies of states $| a \rangle$ and $ | \omega \rangle$, respectively, $ \langle \omega | \omega ' \rangle= \delta (\omega-\omega')$, $ \langle a | a \rangle=1$, $\langle a | \omega \rangle =0$, $g(\omega)$ is the spectral coupling function, and $f(t)$ describes the time-dependent strength of the interaction. We will typically assume a spectral function $g(\omega)$ bounded from below, with $g (\omega_a) \neq 0$, and a  finite interaction time, i.e. we assume $f(t) \rightarrow 0$ as $ t \rightarrow \pm \infty$. The Hamiltonian (1) can describe, for example, photonic transport of a localized mode in an optical waveguide or resonator $|a \rangle$ which is side coupled to a continuum of states $|\omega \rangle$ represented by an array of optical waveguides, a photonic crystal or a slab waveguide \cite{ruffa,ruffa1}.
For the system initially prepared in state $| a \rangle$, the population (photon energy) left in the localized mode $|a \rangle$ after the interaction with the continuum is then defined by $P_{s}= \lim_{t \rightarrow \infty} | \langle a | \psi(t) \rangle|^2$. Here we assume a non-Hermitian coupling by allowing the function $f(t)$ to become {\it complex}, i.e. with a non-vanishing imaginary part. In this case, the total population (photon number) of the system $ \mathcal{N}=|\langle a | \psi(t) \rangle|^2 +\int d \omega | \langle  \omega | \psi(t)\rangle |^2$ is not conserved, because photons can be created or annihilated during the discrete-continuum coupling by the gain/loss in the system that are needed to realize the non-Hermitian coupling (see Sec.4.2 for a detailed discussion on this point). A pseudo decoupling of the discrete state with the continuum is attained whenever $P_s=1$, i.e. after the interaction the discrete state returns to its initial condition, as if the interaction were not occurred at all. While in the Hermitian limit the decoupling condition $P_s=1$ can be realized only approximately and under special conditions, we will show here that decoupling can be obtained for a broad class of {\em complex} coupling functions $f(t)$. However, in the latter case the decoupling is imperfect, i.e. while after the interaction the discrete state returns to its initial condition, as if the interaction had not occurred, some excitation is left in the continuum, which keeps memory of the interaction. Such a pseudo-decoupling is a clear signature of non-Hermitian dynamics and it is prevented for an Hermitian coupling owing to population (photon energy) conservation.  \\
If the state vector $| \psi(t) \rangle$ of the system is decomposed as 
\begin{equation}
| \psi(t) \rangle = \left[ c_a(t) | a \rangle + \int d \omega c(\omega,t) | \omega \rangle \right] \exp(-i \omega_a t)
\end{equation} 
from the equation $ i \partial_t | \psi(t) \rangle = \hat{H} | \psi(t) \rangle$ we obtain the following evolution equations for the modal amplitudes $c_a(t)$ and $c(\omega,t)$ in the discrete and continuum of states
\begin{eqnarray}
i \frac{dc_a}{dt} & = &  f(t) \int d \omega g( \omega) c(\omega,t) \\
i \frac{dc}{dt} & = & (\omega-\omega_a)  c +f(t) g^*(\omega) c_a.
\end{eqnarray}
For a system prepared at initial time $t= -\infty$ in state $| a \rangle$, Eqs.(3) and (4) should be integrated with the initial conditions
\begin{equation}
c_a(-\infty)=1 \; , \;\;\; c(\omega, -\infty)=0
\end{equation}
and the final population of the discrete state  is given by $P_s= \lim_{t \rightarrow \infty} |c_a(t)|^2$.
 \begin{figure}
\includegraphics[scale=0.3]{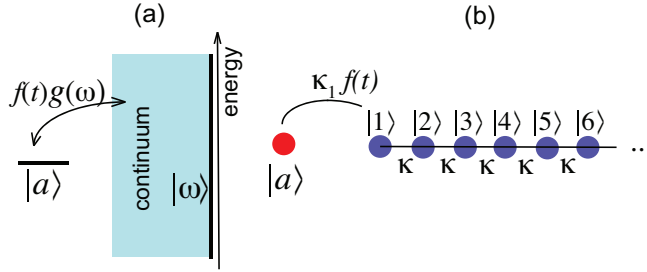}
\caption{(Color online) (a) Schematic of a discrete (bound) state $| a \rangle$ coupled to a continuum of states $| \omega \rangle$ with a time-dependent coupling strength $f(t)$ and spectral function $g(\omega)$. (b) Schematic of the photonic Fano-Anderson model  realized by a site $| a \rangle$ side-coupled to a semi-infinite tight-binding lattice. The sites in the lattice correspond to optical microrings or resonators with evanescent mode coupling.}
\end{figure}
 \begin{figure}
\includegraphics[scale=0.3]{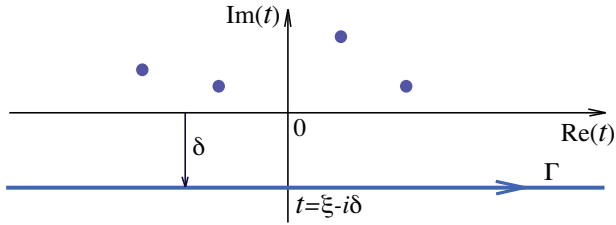}
\caption{(Color online) Schematic showing the integration path of the coupled equations (3) and (4) in complex $t$ plane along the horizontal $\Gamma$ line in the lower half part of the complex plane. The coupling function $f(t)$ is holomorphic in the ${\rm Im}(t) \leq 0$ half plane, with poles in the upper half complex plane (filled circles).}
\end{figure}

\subsection{Weak coupling and adiabatic interaction: decay rate and Fermi golden rule}
To highlight the main features of non-Hermitian coupling versus Hermitian one, let us first restrict our analysis to the weak-coupling (Markovian) limit $f (t) g(\omega) \rightarrow 0$. In this case a generalized Fermi golden rule can be derived following a rather standard procedure which is outlined, for instance, in Ref.\cite{Z10}. After formal integration of Eq.(4) with the initial condition $c(\omega,-\infty)=0$, elimination of the amplitudes $c(\omega,t)$ yields the following exact integro-differential equation for $c_a(t)$
\begin{equation}
\frac{dc_a}{dt}= -f(t) \int_{-\infty}^{t} d t' f(t') \Phi(t-t') c_a(t') 
\end{equation}  
where $\Phi(\tau)$ is the reservoir response (memory) function defined by
\begin{equation}
\Phi(\tau)= \int d \omega |g (\omega)|^2 \exp[-i(\omega-\omega_a) \tau].
\end{equation}
The memory function $\Phi(\tau)$ vanishes as $\tau \rightarrow \infty$ with a characteristic time scale $\tau_m$, the memory time. For a weak coupling $|f(t) g (\omega)| \ll 1 / \tau_m$ and for a coupling function $f(t)$ that varies slowly over the memory time $\tau_m$, Eq.(6) can be approximated with the differential equation
\begin{equation}
\frac{dc_a}{dt} \simeq -f^2(t) (R-i \Delta)  c_a
\end{equation}
where we have set
\begin{equation}
R-i \Delta \equiv \int_{0}^{\infty} d \tau \Phi(\tau),
\end{equation}
  i.e.
  \begin{eqnarray}
  R & = & \pi | g (\omega_a)|^2 \\
  \Delta & = & P  \int \frac{d \omega |g(\omega)|^2}{\omega-\omega_a}
  \end{eqnarray}
  and $P$ denotes the principal value of the integral.
  Integration of Eq.(8) with the initial condition $c_a(-\infty)=1$ yields
  \begin{equation}
  c_a(t)=\exp \left\{ -(R-i \Delta) \int_{-\infty}^{t} dt' f^2(t') \right\}.
  \end{equation}
  Equation (12) provides an extension of the usual exponential decay law in the weak-coupling (Markovian) limit to the case of a 
  time-varying coupling strength $f(t)$, $R$ and $\Delta$ being the ordinary decay rate and frequency shift of the metastable $| a \rangle$ state. The final population in state $|a \rangle$ after the interaction with the continuum reads
\begin{equation}
P_s= \lim_{t \rightarrow \infty} |c_a(t)|^2 = \exp \left\{ -2 {\rm Re} \left( R \mathcal{A}-i \Delta \mathcal{A} \right) \right\}
\end{equation}
where we have set 
\begin{equation}
\mathcal{A} \equiv \int_{-\infty}^{\infty} dt f^2(t)
\end{equation}
 as a definition of the effective interaction time.\\
  In the Hermitian case, $\mathcal{A}$ is real and positive, so that the final population reads
\begin{equation}
P_s=\exp(-2 R \mathcal{A})
\end{equation}
Therefore, assuming that $\omega_a$ is embedded into the continuous spectrum and that decay into the continuum is allowed, i.e. $g(\omega_a) \neq 0$, one always has $P_s<1$, i.e. there is a nonvanishing probability of decay after interaction with the continuum.\\ 
For a non-Hermitian coupling, the function $f(t)$ is complex  
and $P_s$ can be even become larger than one after the interaction. Interestingly, the effective interaction time $\mathcal{A}$ can vanish, indicating that after interaction the state $|a \rangle$ returns to its initial condition despite $\omega_a$ is embedded into the continuous spectrum. Such a counterintuitive result does not depend on the particular form of the spectral coupling $g(\omega)$ and just requires $\mathcal{A}=0$, which can be satisfied by a broad class of complex coupling functions $f(t)$.

\section{Non-Hermitian pseudo decoupling}
In the previous section we have shown that, in the Markovian (weak coupling) limit, a discrete state $ | a \rangle$ interacting with a continuum can return to its initial condition after the interaction whenever the effective time $\mathcal{A}$, defined by Eq.(14), vanishes. While such a possibility is prevented for an Hermitian coupling, it is possible for a broad class of non-Hermitian coupling functions $f(t)$. In this section we prove that, whenever the coupling function $f(t)$  is analytic in the half lower ${\rm Im}(t) \leq 0 $ (or half upper ${\rm Im}(t) \geq 0$) complex plane, the discrete state $|a \rangle$ interacting with the continuum $| \omega \rangle$ can return to its initial state even for strong discrete-continuum coupling and for a coupling function rapidly varying on the time scale of  the memory time $\tau_m$. However, some excitation is generally left in the continuum after the interaction, which thus keeps some memory of the interaction. Therefore, non-Hermitian interaction realizes a discrete-continuum pseudo-decoupling. \\
For the sake of definiteness, we typically assume that the coupling function $f(t)$ is a  meromorphic function, analytic in the half lower ${\rm Im}(t) \leq 0$ complex plane and vanishing as $|t| \rightarrow \infty$. For instance, any function of the form
\begin{equation}
f(t)=\sum_n \frac{A_n}{(t-t_n)^{\alpha_n}}
\end{equation}
satisfies such a requirement,
  where $t_n$ are the poles of $f(t)$ in the upper half complex plane [${\rm Im}(t_n) >0$], $A_n$ are complex amplitudes, and the integer $\alpha_n \geq 1$ is the order of pole $t_n$. Note that, since $f(t)$ is analytic in a half part of the complex plane, the real $f_R(t)$ and imaginary $f_I(t)$ parts of $f(t)$ are related each other by a Hilbert transform, i.e. by Kramers-Kronig relations.\\ 
  At initial time $t \rightarrow -\infty$, we assume $c_a(-\infty)=1$ and $c(\omega,-\infty)=0$, corresponding to excitation of the discrete state and to an empty continuum. Since $f(t)$ is analytic in the half plane ${\rm Im}(t) \leq 0$, Eqs.(3) and (4) can be extended into the complex $t$ plane and the solution $c_a(t)$ and $c(\omega,t)$ can be analytically continued in the ${\rm Im}(t) \leq 0$ half plane. We wish to prove that $c_a(\infty)=1$, i.e. after the interaction the discrete state returns to its initial condition, corresponding to an effective decoupling to the continuum. However, rather generally one has $c(\omega, \infty) \neq 0$, indicating that the discrete-continnum decoupling is asymmetric and memory of the interaction is kept in the continuum. To this aim, let us integrate Eqs.(3) and (4) along the horizontal line $\Gamma$ of the complex plane shown in Fig.2, $t= \xi-i\delta$ with $\delta>0$ and $-\infty < \xi < \infty$. The solution will depend parametrically on $\delta$. Since $f(\xi-i \delta)$ vanishes as $ \xi \rightarrow \pm \infty$, the solutions $c_a (\delta, \xi)$ and $c(\omega,\delta, \xi)$ to Eqs.(3) and (4) along the line $\Gamma$, corresponding to occupation of the discrete state solely and to an empty continuum at $\xi \rightarrow -\infty$, have the asymptotic behavior
  \begin{equation}
  c_a(\delta, \xi) \sim 
  \left\{
  \begin{array}{ll}
  1 & \xi \rightarrow - \infty \\
  A(\delta)  & \xi \rightarrow \infty
  \end{array}
  \right. 
  \end{equation}
  \begin{equation}
  c(\delta, \omega, \xi) \sim 
  \left\{
  \begin{array}{ll}
  0 & \xi \rightarrow - \infty \\
  B (\delta, \omega) \exp[-i (\omega-\omega_a) \xi ] & \xi \rightarrow \infty
  \end{array}
  \right. 
  \end{equation}
  where $A(\delta)$ and $B(\delta, \omega)$ are integration constants, independent of $\xi$.  
Note that the population in $|a \rangle$ after the interaction with the continuum is given by 
\begin{equation}
P_s= \lim_{\xi \rightarrow \infty} |c_a(\delta=0, \xi)|^2=|A(\delta=0)|^2.
\end{equation}
Since $c_a(\delta,\xi)$ is an analytic function of $t= \xi-i \delta$ in a stripe that encloses the line $\Gamma$, one has $(\partial c_a / \partial \delta)=-i (\partial c_a / \partial\xi)$, which from Eq.(17) and taking the limit $ \xi \rightarrow \infty$ yields $ (dA/ d \delta)=0$, i.e.
  \begin{equation}
  A(\delta)=A(0).
  \end{equation}
 Therefore, to calculate the final population $P_s$ we can integrate Eqs.(3) and (4) along any arbitrary horizontal line $\Gamma$ in the lower half complex plane, and in particular by taking $\delta$ arbitrarily large. After elimination of $c(\delta,\omega,t)$, from Eqs.(3) and (4) the exact integro-differential equation for $c_a(\delta, \xi)$ is readily obtained
\begin{equation}
\frac{\partial c_a}{\partial \xi}=- f(\xi-i \delta)  \int_{-\infty}^{\xi} d \theta f(\theta -i \delta) \Phi(\xi-\theta) c_a(\delta,\theta)
\end{equation}
where $\Phi(\tau)$ is the memory function defined by Eq.(7). Note that Eq.(21) is analogous to Eq.(6), but the solution is now computed along the line $\Gamma$. For a large value of $\delta$, $f(\xi-i \delta)$ becomes arbitrarily small together with its derivative with respect to $\xi$. Hence for large $\delta$ we can replace Eq.(21) with the differential equation
\begin{equation}
\frac{\partial c_a}{\partial \xi}=- f^2(\xi-i \delta)(R-i \Delta) c_a(\delta,\xi).
\end{equation}
Note that, while Eq.(22) is only an approximate result at $\delta=0$ and corresponds to the Markovian approximation discussed in Sec.II.B, it is an exact result in the $\delta \rightarrow \infty$ limit, i.e. when Eqs.(3) and (4) are integrated in the complex plane along the line $\Gamma$ far below the real axis. Integration of Eq.(22) yields
\begin{equation}
c_a(\delta,\xi)=\exp \left\{ -(R-i \Delta) \int_{-\infty}^{\xi} d \theta f^2(\theta-i \delta)  \right\}
\end{equation}
and thus
\begin{equation}
A(\delta)= \lim_{\xi \rightarrow \infty} c_a( \delta, \xi)=1
\end{equation}
where we used the property
\begin{equation}
\int_{-\infty}^{\infty} d \theta f^2(\theta-i \delta)=0
\end{equation}
that follows from the application of the residue theorem. Therefore, after the interaction $c_a(\infty)=1$ as if the interaction with the continuum had not occured. However,  some excitation is generally left in the continuum after the interaction, i.e. $c(\omega, \infty) \neq 0$ corresponding to an increase of the norm of the discrete-continuum system after the interaction. In fact, the analyticity condition of $c(\omega,t=\xi-i \delta)$ yields $(\partial c  / \partial \delta)=-i (\partial c / \partial\xi)$, and thus from Eq.(18) it follows that the amplitude $B(\delta, \omega)$ is found to vary with $\delta$ as $B(\delta,\omega)=B(0,\omega) \exp[(\omega-\omega_a) \delta ]$ [compare with Eq.(20)]. The excitation left in the continuum after the interaction can be thus computed as $B(0,\omega)= \lim_{\delta \rightarrow \infty }B(\delta,\omega) \exp[-(\omega-\omega_a) \delta ]$. Since $B(\delta,\omega) \rightarrow 0$ as $\delta \rightarrow \infty$, one obtains $B(0,\omega)=0$ for $\omega \geq  \omega_a$. However, for $\omega<\omega_a$ an indetermination form $0 \times \infty$ is found, so that a nonvanishing amplitude $B(0,\omega)$ can be found for the spectral components $\omega<\omega_a$ of the continuum after the interaction.
 Such a result shows that the continuum keeps memory of the interaction and a pseudo decoupling is observed: while the interaction does not have any measurable effect on the discrete state, it leaves some excitation into the continuum which was initially empty.\par
The above analysis suggests a rather general strategy of pseudo decoupling based on non-Hermitian dynamics. For a decaying state $|a \rangle$  interacting with a continuum via an Hermitian coupling described by a {\em real}  function $f_R(t)$, an effective decay suppression can be realized by adding a non-Hermitian coupling so as the interaction is described by the {\it complex} function $f(t)=f_R(t)+i f_I(t)$, where $f_I(t)$ is the Hilbert transform of $f_R(t)$.

 \begin{figure}
\includegraphics[scale=0.3]{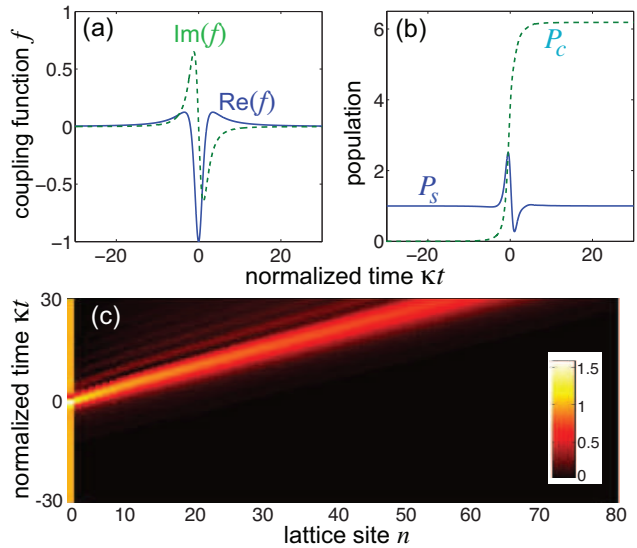}
\caption{(Color online) Pseudo decoupling of a discrete state $|a \rangle$ side-coupled to a semi-infinite tight-binding lattice via a non-Hermitian interaction. (a) Temporal behavior of the coupling function $f(t)$. The real and imaginary parts of $f$ are related each other by a Hilbert transform. (b) Numerically-computed behavior of the populations  $P_s(t)$ and $P_c(t)$ of the discrete and continuum states. Note that the total population (photon energy) $P_s(t)+P_c(t)$ is not conserved owing to non-Hermitian interaction. (c) Snapshot, on a pseudo color map, of the temporal evolution of the occupation amplitudes $|c_a(t)|$, $|c_n(t)|$. The site $n=0$ corresponds to the edge site $| a \rangle$. Parameter values are given in the text.}
\end{figure}

 \begin{figure}
\includegraphics[scale=0.3]{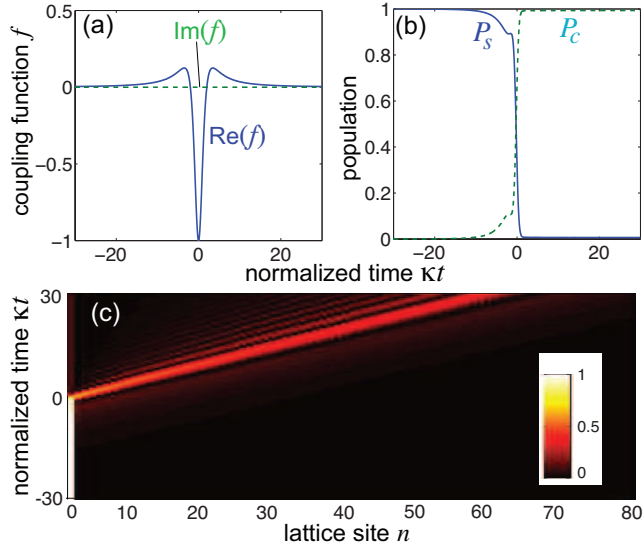}
\caption{(Color online) Same as Fig.3, but for a real coupling function $f(t)$ (Hermitian coupling).}
\end{figure}

\section{An example: tight-binding Fano-Anderson model with effective non-Hermitian coupling}
\subsection{Non-Hermitian Fano-Anderson model}
In this section we illustrate the concept of non-Hermitian pseudo decoupling by considering a paradigmatic model of discrete-continuum interaction, namely the Fano-Anderson model \cite{F1,F2} realized in a tight-binding lattice. The Hermitian Fano-Anderson model has been widely encountered and studied in different physical contexts \cite{F3A,F3B,F3C,F4A,F4B,F5A,F5B,F5C,F5D,F6A,F6B,F6C,F6D,F6E,F6F}. In mesoscopic systems it describes, for example, the decay dynamics of a quantum dot coupled to a quantum wire, whereas in optics it models the escape dynamics of light waves in evanescently-coupled optical waveguide arrays or coupled optical resonators.\\ 
Let us consider the tight-binding lattice shown in Fig.1(b) in which an edge site $|a \rangle$ is coupled to a semi-infinite chain via a non-Hermitian and time-varying coupling $\kappa_1 f(t)$. A physical  implementation of non-Hermitian coupling for a photonic structure will be discussed in the next subsection. The tight-binding Hamiltonian of the system reads
\begin{eqnarray}
\hat{H} & = &  \omega_a | a \rangle \langle a | - \kappa \sum_{n=1}^{\infty}  \left( | n \rangle \langle n+1|+ | n+1 \rangle \langle n | \right) \nonumber \\
& - & \kappa_1 f(t) \left( |a \rangle \langle 1 | + |1 \rangle \langle a | \right)
\end{eqnarray}
where $\kappa$ is the hopping amplitude between adjacent sites $|n \rangle$ in the chain,  $\omega_a$ is the energy offset of the edge site $| a \rangle$ from the center of the tight-binding lattice band and $f(t)$ is the complex coupling function. In the Bloch function basis $|k \rangle$, defined by
\begin{equation}
| k \rangle = - \sqrt{\frac{2}{\pi}} \sum_{n=1}^{\infty} \sin (nk) | n \rangle 
\end{equation}
the Hamiltonian (26) can be cast in the form \cite{F5A}
\begin{eqnarray}
\hat{H} & = & \omega_a | a \rangle \langle a |+ \int_0^{\pi} dk \omega(k) |k \rangle \langle k|  \nonumber \\
& + & \int_0^{\pi} dk f(t) V(k) \left( |a \rangle \langle k| + |k \rangle \langle a|  \right) 
\end{eqnarray}
where we have set
\begin{eqnarray}
\omega(k) & = & - 2 \kappa \cos (k) \\
V(k) & = & \sqrt{\frac{2}{\pi}} \kappa_1 \sin (k)
\end{eqnarray}
 Note that, after setting $| \omega \rangle= (dk/d \omega)^{1/2} | k \rangle$, the Hamiltonian (28) is of the form (1) with the spectral coupling function $v(\omega)$ given by
 \begin{equation}
 v(\omega)=\sqrt{\frac{dk}{d \omega}} V(k(\omega))=\sqrt{\frac{1}{2 \pi}} \frac{\kappa_1}{\kappa_0} (4 \kappa^2-\omega^2)^{1/4}.
 \end{equation}
 If the state vector of the system is expanded on the basis of Wannier states $| n \rangle$, i.e. after setting
 \begin{equation}
 | \psi(t) \rangle = c_a(t) |a \rangle +\sum_{n=1}^{\infty} c_n(t) | n \rangle
 \end{equation}
 the evolution equations of the amplitudes $c_a$ and $c_n$ read
 \begin{eqnarray}
 i \frac{dc_a}{dt} & = & \omega_a c_a -\kappa_1 f(t) c_1 \\
 i \frac{dc_1}{dt} & = & -\kappa_1 f(t) c_a- \kappa c_2 \\
 i \frac{dc_n}{dt} & = &  -\kappa (c_{n+1}+c_{n-1} )\;\;\; (n \geq 2) 
 \end{eqnarray}
 We checked the possibility of pseudo decoupling for a wide class of complex functions $f(t)$, such as the class of meromorphic functions [Eq.(16)], by direct numerical simulations of Eqs.(33-35). As an example, Fig.3 shows the numerically-computed results obtained for a meromorphic function with a second-order pole, $f(t)=1/ (\kappa t-t_1)^2$ [Fig.3(a)], for parameter values $\kappa_1 / \kappa=2$, $\omega_a / \kappa=0$ and $t_1 =2i $. Figure 3(b) shows the numerically-computed behavior of the population $P_s(t)=|\langle a | \psi(t) \rangle|^2=|c_a(t)|^2$ at site $|a \rangle$ and of the population $P_c(t)$ in the continuum (the other sites in the chain), $P_c(t)=\sum_{n=1}^{\infty} |c_n(t)|^2$. Note that, since the Hamiltonian in not Hermitian, the total population $P_s(t)+P_c(t)$ is not conserved and can even become larger than its initial value owing to gain/loss in the system. Figure 3(c) shows on a pseudocolor map the detailed evolution of the amplitudes in Wannier basis, i.e. $|c_n(t)|$ ($n=1,2,3,...$) and $|c_a(t)|$ (the site $n=0$ in the figure). Figures 3(b) and (c)  clearly show that, after the interaction of the site $|a \rangle$ with the continuum at around $t=0$, the site $| a \rangle$ is not affected by the interaction, as if the interaction had not occured. However, the continuum keeps memory of the interaction, since some excitation in the continuum after the interaction is observed (clearly indicated by the non-vanishing value of $P_c(t)$ after the interaction). For comparison, Fig.4 shows the numerical results of discrete-continuum interaction when the coupling function $f(t)$ is assumed to be real [we disregarded  the imaginary part of the meromorphic function $f(t)$]. In this case, corresponding to Hermitian interaction and conservation of the full population $P_s(t)+P_c(t)=1$, almost complete decay of the discrete state into the continuum is clearly observed. 
 
\subsection{Physical implementation}
A physical implementation of a non-Hermitian (complex) hopping rate between site $|a \rangle$ and the tight-binding chain $\{ |n \rangle \}$ [Fig.1(b)] can be obtained by considering a fast modulation of the site energy potential in $|a \rangle$ with complex (i.e. loss and gain) values. The tight-binding chain with modulated term of the edge site $| a \rangle$ is described by the Hamiltonian
\begin{eqnarray}
\hat{H} & = &  [\omega_a+A(t) \cos (\Omega t)] | a \rangle \langle a | \nonumber \\
& -&  \kappa \sum_{n=1}^{\infty}  \left( | n \rangle \langle n+1|+ | n+1 \rangle \langle n | \right) \nonumber \\& - & \kappa_1 \left( |a \rangle \langle 1 | + |1 \rangle \langle a | \right)
\end{eqnarray}
 \begin{figure}
\includegraphics[scale=0.3]{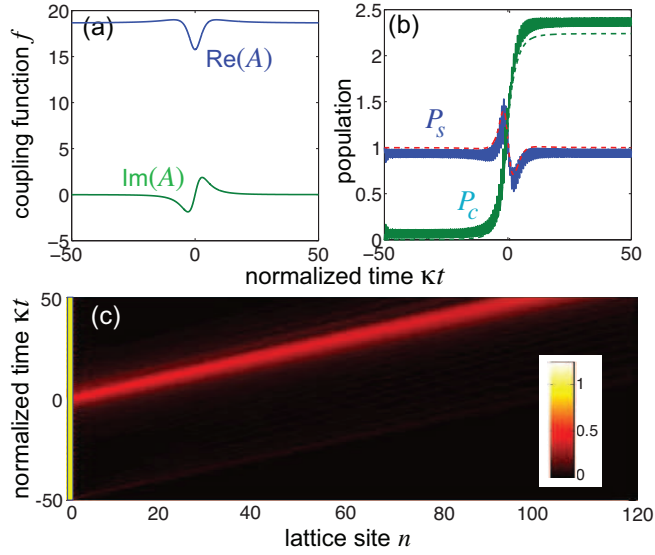}
\caption{(Color online) Pseudo decoupling of a discrete state $|a \rangle$ interacting with a tight-binding lattice via an effective non-Hermitian coupling, implemented by high-frequency on-site potential modulation. (a) Temporal behavior of the modulation envelope $A(t)$ (real and imaginary parts). (b) Numerically-computed behavior of the populations $P_s(t)$ and $P_c(t)$ in the discrete and continuum states. The dashed curves show the behaviors of $P_s$ and $P_c$ as predicted within the effective Hamiltonian (37) under the rotating-wave approximation. (c) Snapshot, on a pseudo color map, of the temporal evolution of the occupation amplitudes  $|c_a(t)|$, $|c_n(t)|$. The site $n=0$ corresponds to the edge site $| a \rangle$. Parameter values are given in the text.}
\end{figure}

 \begin{figure}
\includegraphics[scale=0.3]{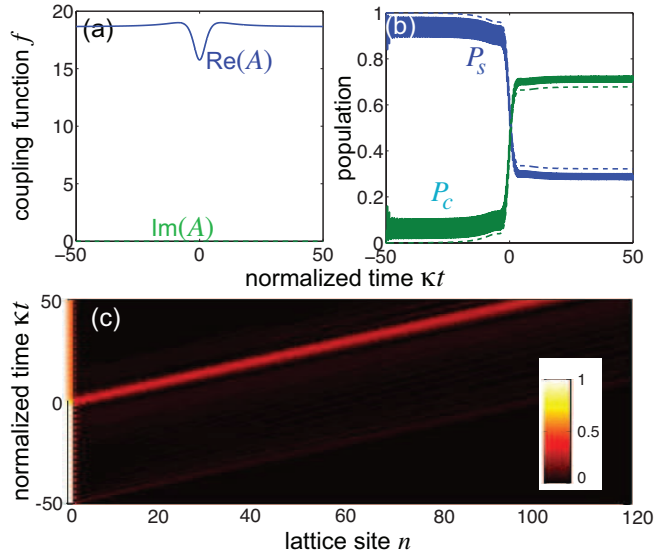}
\caption{(Color online) Same as Fig.5, but for a real modulation function $A(t)$ (effective Hermitian coupling).}
\end{figure}

where $\kappa$ is the hopping amplitude between adjacent sites $|n \rangle$ in the chain,  $\kappa_1$ is the (Hermitian) hopping rate between sites $|a \rangle$ and $| 1 \rangle$, $\omega_a$ is the frequency detuning of site $| a \rangle$, and $A(t)$ is the slowly-varying complex amplitude of the rapidly oscillating part of potential in $|a \rangle$ (oscillation frequency $\Omega$). The tight-binding model (36) with complex on-site oscillating potential can be realized, for example, by considering a chain of coupled microring resonators, in which the oscillating complex potential at the edge site is obtained by combined index and loss/gain modulation of the boundary resonator \cite{resonatorA,resonatorB,resonatorC,resonatorD,resonatorE,resonatorF}. For a fast oscillation, $\Omega \gg \kappa_1,  | \omega_a |$, the rapidly-oscillating term can be averaged after a gauge transformation \cite{rotating}, and the Hamiltonian (36) can be replaced by the effective Hamiltonian (rotating-wave approximation)
\begin{eqnarray}
\hat{H}_{eff} & = &  \omega_a | a \rangle \langle a | - \kappa \sum_{n=1}^{\infty}  \left( | n \rangle \langle n+1|+ | n+1 \rangle \langle n | \right) \nonumber \\
& - & \kappa_1 f(t) \left( |a \rangle \langle 1 | + |1 \rangle \langle a | \right)
\end{eqnarray}
where we have set
\begin{equation}
f(t)=J_0 \left ( A(t) / \Omega \right)
\end{equation}
and $J_0$ is the Bessel function of first kind and zero order. Note that, by taking 
\begin{equation}
A(t)= \Omega[A_0+ \delta A(t)]
\end{equation}
 with $A_0 \simeq 2.405$ (first zero of Bessel $J_0$ function) and $\delta A(t)$ vanishing as $ t \rightarrow \pm \infty$, coherent destruction of tunneling \cite{CDTA,CDTB,CDTC}, i.e. absence of interaction, is attained at $ t \rightarrow \pm \infty$. For small $|\delta A(t)|$, after Taylor expansion one can also write
\begin{equation}
f(t) \simeq J_0'(0) \delta A(t) \simeq -0.5191 \times \delta A (t)
\end{equation}
indicating that, by proper tailoring the slowly-varying amplitude $A(t)$ of the fast oscillation, one can implement an effective non-Hermitian hopping amplitude according to Eq.(40). We checked the feasibility of the method by direct numerical integration of the periodically-driven Hamiltonian (36) without the rotating-wave approximation. Figure 5 shows an example of non-Hermitian pseudo decoupling for parameter values $\omega_a / \kappa=0$, $\kappa_1 / \kappa=2$, $\Omega / \kappa=8$ and for a modulation envelope $A(t)$ defined by Eq.(39) with $A_0=2.33$ and $\delta A (t)=3 /(\kappa t-5 i)^2$. The value of $A_0$, which slightly deviates from the first root of the Bessel $J_0$ function, is chosen so that to decouple the state $| a \rangle$ from the chain $\{ | n \rangle \}$ at $t \rightarrow \pm \infty$ via coherent destruction of tunneling \cite{rotating} \footnote{The decoupling in the Hermitian case at the special amplitude $A(t)=A_0$ corresponds to the existence of a Floquet bound state in the continuum [S. Longhi and G. Della Valle, Sci. Rep. {\bf 3}, 2219 (2013)].}. The behavior of the population  $P_s(t)$ in the discrete state and of the population in the continuum $P_c(t)$, as predicted by the effective Hamiltonian (37) under the rotating-wave approximation, is also shown for comparison by dashed curves in Fig.5(b). The numerical results corresponding to an Hermitian coupling, obtained by setting equal to zero the imaginary part of $\delta A(t)$, are shown in Fig.6. In this case after interaction with the continuum there is clearly a non-vanishing probability of decay of state $|a \rangle$.

\section{Conclusion}
Controlling the dynamics of a discrete state coupled to a continuum is a subject of major relevance in different areas of quantum and classical physics. 
In Hermitian systems, several decoupling methods have been suggested, where decoupling results in the suppression of transitions between the discrete state and the continuum. In open quantum or classical systems, however, the discrete-continuum dynamics can become non-unitary and described by an effective non-Hermitian Hamiltonian. For example, in photonic transport in materials with optical gain and loss the total number of photons is not conserved, and the evanescent coupling of a localized mode with a continuum of states can be engineered to realize a kind of non-Hermitian interaction. 
 In this work we have theoretically investigated the dynamics of a discrete state coupled to a continuum of states via a time-dependent {\em non-Hermitian interaction} and have  disclosed a kind of pseudo dynamical decoupling, which does not have any counterpart in Hermitian systems. For a given time-dependent Hermitian coupling $f_R(t)$, that would result in the decay of the discrete state into the continuum, the addition of an imaginary  (non-Hermitian) coupling $f_I(t)$, such that $f_R(t)$ and $f_I(t)$ are related each other by a Hilbert transform, leaves the discrete state in its initial condition, as if the interaction had not occurred. However, trace of the interaction is kept in the continuum, with a non-vanishing excitation of the continuum after its interaction with the discrete state. The concept of non-Hermitian pseudo decoupling has been exemplified by considering a non-Hermitian extension of the Fano-Anderson model for photonic transport in a tight-binding lattice. Our results  provide novel insights into the non-Hermitian properties of a discrete state coupled to a continuum of states and disclose important dynamical features with no counterparts in ordinary Hermitian systems, such as the possibility to realize a vanishing interaction time and pseudo decoupling of the discrete-continuum states.

\end{document}